\documentclass{ws-ijmpa}
\usepackage[super,compress]{cite}
\usepackage{graphicx}

\usepackage{soul}
\usepackage{xcolor}

\begin{document}
\markboth{Boris Tom\'a\v{s}ik}{On elliptic flow and the blast-wave model}

%
\catchline{}{}{}{}{}
%

\title{On elliptic flow and the blast-wave model\footnote{Minireview for the Zim\'anyi School volume}
}

\author{Boris Tom\'a\v{s}ik}

\address{Fakulta jadern\'a a fyzik\'aln\v{e} in\v{z}en\'yrsk\'a, \v{C}esk\'e vysok\'e u\v{c}en\'i technick\'e v Praze, 
B\v{r}ehov\'a 7,\\ 11519 Praha 1, Czech Republic\\ and\\
Fakulta pr\'irodn\'ych vied, Univerzita Mateja Bela, Tajovsk\'eho 40, 97401 Bansk\'a Bystrica, Slovakia\\
boris.tomasik@cvut.cz}

\maketitle

\begin{history}
\end{history}

\begin{abstract}
Extensions of the blast-wave model for the description of non-central collisions are reviewed and the 
physics behind them is explained and illustrated. It is shown how the second-order anisotropy in expansion 
velocity together with the anisotropy in the shape or the density profile of the fireball determine the 
elliptic flow of the produced hadrons. Ambiguities and limitations of the models are discussed and different 
models are compared among themselves and with example data. It is concluded that model results should always 
be compared to $v_2(p_t)$ from several identified species in order to receive meaningful results about 
the freeze-out stage of the fireball and that even such a comparison may not be conclusive as the models 
may not be able to reproduce all relevant data. 

\end{abstract}



\section{Introduction}	

Elliptic flow belongs to the most prominent observables in physics of ultrarelativistic heavy-ion collisions \cite{Danielewicz:1985hn,Ollitrault:1992bk,Voloshin:1994mz,Huovinen:2001cy}.
Through a chain of implications it provides access to the medium response to high energy density, i.e., the actual Equation of State and transport properties of the hot strongly interacting matter. 
Such a logical link is implicitly implemented in a simulation that uses  hydrodynamic model. 
Nevertheless, in order to get an economic, yet still physically motivated description of data, fits with model parametrizations of hadron emission at freeze-out based on  some underlying physics picture are often performed. 
A prominent place among such parametrizations is occupied  by the blast-wave model \cite{Siemens:1978pb,Schnedermann:1993ws,Csorgo:1995bi,Tomasik:1999cq,Retiere:2003kf,Tomasik:2004bn,Cimerman:2017lmm}.

The following notes summarize  the interpretation, use, and abuse of the model in reproducing the elliptic flow. 
My aim is to put together ideas and formulas that are scattered in the literature and explain their physics motivation. 
I hope that this may be useful for those students who will try to understand fits with such formulas at conferences or in papers. 
Also, I would like to make it useful for those, who will use such formulas in data analyses. 

Two things are implicitly addressed by the name ``elliptic flow''. 
On an observational level, distribution of the produced hadrons shows an anisotropy in the azimuthal plane, mainly in non-central collisions. 
Its dominant component is usually elliptic: more hadrons are produced within the so-called event plane\footnote{%
The second-order event plane is spanned by the beam direction and the direction of the vector 
$\vec Q = \sum_i p_{t,i} (\cos(2 \phi_i),\sin(2\phi_i))$, where $p_{t,i}$ 
and $\phi_i$ are transverse momentum and azimuthal angle of the $i$-th particle, and the sum goes over all particles in an event.
}
than perpendicularly to it. 
The event plane is a measurable proxy for the reaction plane, which is given by the direction of the beam and the impact parameter. 
Within hydrodynamics, the interpretation of the anisotropic distribution is at hand: it results from an anisotropic expansion of the hot matter that is faster within the reaction plane than perpendicularly to it. 
This is the second meaning of the label ``elliptic flow'', and it is why it refers to \emph{flow}. 
Such a flow anisotropy is a consequence of higher acceleration within the reaction plane, due to higher initial pressure gradients in this direction, as resulting from the collision geometry. 
In this way, the observable anisotropy of hadron distributions is connected to pressure and the equation of state. 

The second-order anisotropy is quantified by the Fourier coefficient of the azimuthal distribution of hadrons\cite{Voloshin:1994mz}
\begin{equation}
v_2(p_t,y) = \frac{\int_0^{2\pi} d\psi\, \frac{d^3N}{p_t\, dp_t\, dy\, d\psi}\cos(2\psi)}{\int_0^{2\pi} d\psi\, \frac{d^3N}{p_t\, dp_t\, dy\, d\psi}}\, ,
\label{e:1}
\end{equation}
where we assume that the event plane coincides with the $\psi=0$ direction.
The coefficient $v_2$ depends on $p_t$ of the hadrons---which will be explored in these notes---and the rapidity $y$. 
We disregard the latter dependence;  it should be constant in collisions at  very high energies due to longitudinal boost invariance.

In the limit $p_t\to 0$, coefficient $v_2(p_t)$ must vanish, because otherwise the distribution of produced hadrons
in momenta would  be discontinuous when both $p_x, p_y \to 0$.
The size and growth of $v_2(p_t)$ at non-zero $p_t$ reflects the underlying anisotropies within the fireball.

The elliptic flow is one of the basic observables that are usually calculated in hydrodynamic, transport, or hybrid simulations of heavy-ion collisions.
Parameters of the models are nowadays tuned in extensive Bayesian analyses in which the models are compared to data.
Nevertheless, in order to get a first quick, but physically motivated insight into the message that experimental data convey, formulas motivated by blast-wave model are used to fit them.

Unfortunately, the formulation of the blast-wave model for non-central collision events is not uniquely established in literature, and some of the models
that exist even have poor physics motivation.
Yet more seriously, it may actually be so distant from reality that the validity of any quantitative results obtained from such a fit is questionable.

To explain my points, and first of all to set the stage for the discussion, in the next Section I will introduce two possible extensions of the blast-wave model that can be traced in literature.
Then, in Section \ref{s:exams}
I will show some resulting $v_2(p_t)$ for various realistic settings of the models and even do a very simple comparison to a small set of data.
Based on these findings, conclusions will be formulated in Section \ref{s:conc}.


\section{The Blast-Wave Model}
\label{s:model}

The idea of the model is that it describes \emph{sudden} kinetic freeze-out of the whole expanding fireball. 
This is a dramatic approximation motivated by the aim to simplify the description of the freeze-out so that it is reasonably tractable, to some extent even analytically. 
Note that the model assumes that the freeze-out is common for all hadron species, hence production of them all should be described by the same parameters, like temperature and transverse expansion velocity. 
Even if one would be willing to give up such an assumption in order to mimic gradual decoupling of different species, the 
minimum assumption would be that pions and nucleons freeze-out simultaneously, as the former mediate the strong interaction of the latter. 

The model has evolved historically and currently has well established azimuthally symmetric version, even though small variations appear across the literature. 
Such version is applicable to ultra-central collisions or for an effective description of azimuthally-integrated data.

\subsection{The Basics}

The blast-wave model is based on the Cooper-Frye formula\cite{Cooper:1974mv}
\begin{equation}
\label{e:CF}
E\frac{d^3N}{dp^3} = \int f(x,p) \, p_\mu d^3\Sigma^\mu = \int S(x,p) d^4x  .
\end{equation}
This assumes freeze-out along three-dimensional hypersurface $\Sigma$, that divides the space-time into the region where particles interact and the region where they freely stream towards detector. 
The infinitesimal surface element vector $d^3\Sigma^\mu$ points into the free-streaming region and $p_\mu d^3\Sigma^\mu$ gives the flux of  the decoupling hadrons through the hypersurface. 
Note that the orientation of $d^3\Sigma^\mu$ can in general be
such that the hypersurface is space-like as well as time-like,
though in this study---and in the blast-wave model in general---only space-like hypersurfaces 
will be investigated.
 Next, $f(x,p)$ is the local distribution of hadrons that undergo the freeze-out. 
In the second integral, the distribution $f(x,p)$ and a part of the flux factor are included into the \emph{emission function} $S(x,p)$. 

The vanilla blast-wave model that will be used here is given by the emission function
\begin{equation}
\label{e:Sbase}
S(x,p) d^4x = \frac{1}{(2\pi)^3} \frac{1}{\exp\left ( \frac{p_\nu u^\nu}{T} \right ) \pm 1} \theta(R_0-r) \delta(\tau - \tau_0) \, m_t \cosh(y-\eta)\, 
d\tau\, \tau \, d\eta\, r\, dr\, d\phi\,  ,
\end{equation}
where 
\begin{itemize}
\item An assumption of  longitudinal boost-invariance \cite{Bjorken:1982qr} makes the hyperbolic coordinates particularly appropriate: 
$$
\eta  = \frac{1}{2} \ln \frac{t+z}{t-z}\, , \qquad \tau = \sqrt{t^2 - z^2} \, .
$$
Polar coordinates $r$ and $\phi$ in the plane transverse to the beam are chosen as they are suitable due to  azimuthal symmetry. 
\item In the transverse plane, the fireball has circular profile  with radius $R_0$.
\item Freeze-out happens along the hyperbola fixed by $\tau =  \tau_0$, where $\tau_0$ is model parameter.  The time of the freeze-out
does not depend on $r$. This hypersurface is illustrated in Fig.~\ref{f:fohs}.
\begin{figure}[t]
\centerline{\includegraphics[width=0.51\textwidth]{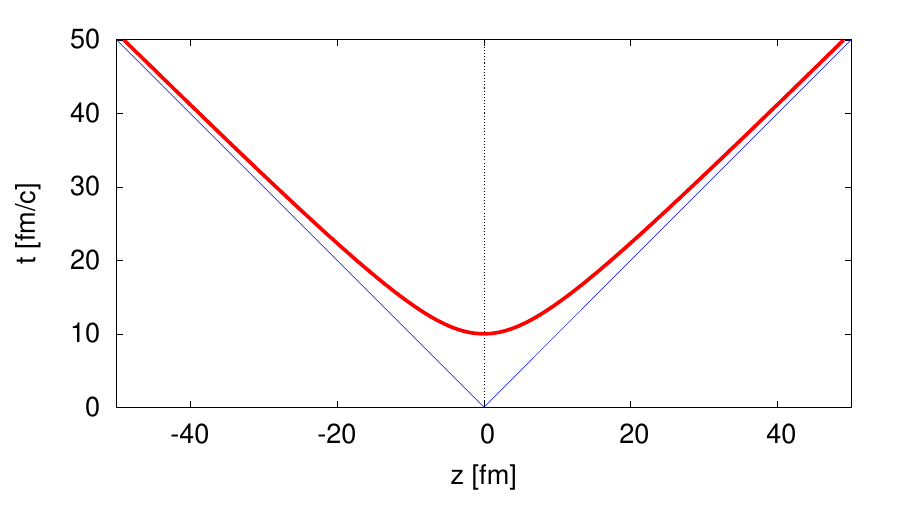}
\includegraphics[width=0.3\textwidth]{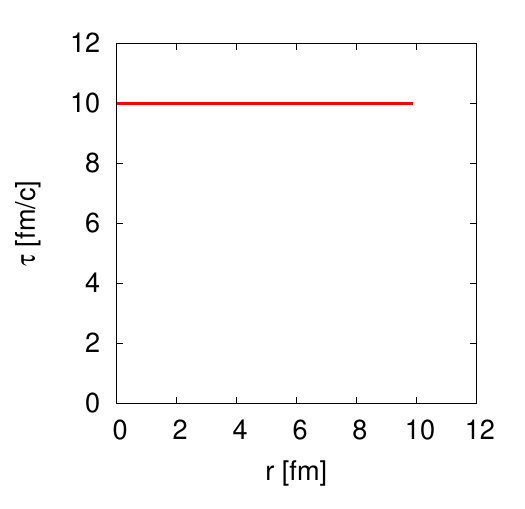}}
\caption{The freeze-out hypersurface for $\tau_0 = 10$~fm/$c$. Left: the profile in the $zt$ plane. Right: the profile in the $\tau r$ plane.
\label{f:fohs}
}
\end{figure}
%
\item 
The expansion of the fireball fluid is parametrized by the velocity field
\begin{eqnarray}
u^\mu & = & \left (
\cosh\eta\, \cosh\eta_t(r),\, \sinh\eta_t(r)\, \cos\phi,\,  \right .
\nonumber
\\
& & 
\left . 
\qquad \qquad \sinh\eta_t(r)\,\sin\phi,\, \sinh\eta\,\cosh\eta_t(r)
\right )\, ,
\\
\eta_t(r) & = & \rho_0 \frac{r}{R_0} = \rho_0 \tilde r\, ,
\end{eqnarray}
which also defines the dimensionless radial coordinate $\tilde r$, and $\rho_0$ is the model parameter measuring the strength of the transverse flow.
These relations define a longitudinally boost-invariant flow field\cite{Bjorken:1982qr}. Transverse velocity (in the frame that co-moves longitudinally 
with the fluid) is obtained from the transverse rapidity $\eta_t(r)$ as 
\[
v_t(r) = \tanh\eta_t(r)\, .
\]
\item 
The distribution of hadrons just before the freeze-out is given by Fermi-Dirac or Bose-Einstein distribution, and $p_\nu u^\nu$ gives the energy of the hadron in the local rest frame of the expanding fluid. 
\item 
The spatial distribution of the hadrons is uniform within the transverse radius of the fireball. This is expressed by the $\theta$-function. 
\end{itemize}
When considering the generalisation of the emission function (\ref{e:Sbase}) for non-central collisions which break azimuthal symmetry, there are two principal issues that should be addressed:
\begin{enumerate}
\item The transverse flow velocity will vary with azimuthal angle, in addition to already applied dependence on $r$. In general, the length of the velocity vector can be varied as well as its orientation, which may differ from the radial direction. 
\item The transverse spatial distribution of the just-emitted hadrons---which was flat within circular profile so far---will depend on the azimuthal angle. 
\end{enumerate}
I will elaborate on both these points below and in  subsections \ref{ss:STAR} and \ref{s:sani}.

Let us start with the first feature, which is shared by all models that will be looked upon here. Transverse rapidity will be varied\footnote{%
Note that a slightly different variation can be also found in the literature: $\eta_t = \tilde r ( \rho_0 + 2\rho_2\cos\phi_b)$.
I did not choose it here, because in my version the $\rho_2$ is \emph{relative} amplitude of the variation, analogically to $v_2$ which 
gives \emph{relative} azimuthal variation of the single-particle spectrum.
} as
\begin{equation}
\label{e:asvelo}
\eta_t = \tilde r \rho_0 ( 1 + 2\rho_2\cos\phi_b)\, .
\end{equation}
Since velocity is a vector, we must also specify its direction. 
It turns out from the azimuthal dependence of femtoscopic radii\cite{Tomasik:2004bn,ALICE:2017gxt} 
that the flow appears rather directed  perpendicularly to the surface of the fireball. 
This  is denoted by the angle $\phi_b = \phi_b(\phi)$, which may be different from $\phi$ in case of a deformed fireball (see below). 
This angle has already been introduced in the parametrisation (\ref{e:asvelo}).  The transverse velocity components are then 
\begin{subequations}
\label{e:asveloc}
\begin{eqnarray}
v_x & = & \tanh\eta_t(\tilde r,\phi)\, \cos\phi_b\\
v_y & = & \tanh\eta_t(\tilde r,\phi)\, \sin\phi_b\,  .
\end{eqnarray}
\end{subequations}

\subsection{Observables}

The single-particle spectrum is obtained by inserting the emission function (\ref{e:Sbase}) into Eq.~(\ref{e:CF}). 
After performing the $\tau$-integration, this leads to
\begin{equation}
E \frac{d^3N}{dp^3} = \frac{d^3N}{p_t\, dp_t\, d\psi\, dy}
=   \frac{1}{(2\pi)^3} \int_0^1 d\tilde r \, \tilde r\, R_0^2 \int_0^{2\pi} d\phi \, \int_{-\infty}^{\infty} d\eta 
 \frac{\tau_0\,  m_t \,\cosh(y-\eta)}{\exp\left ( \frac{p_\nu u^\nu}{T} \right ) \pm 1}  \, .
\end{equation}
Quantum-statistical distributions can be re-cast as infinite series
\begin{equation}
\frac{1}{\exp\left ( \frac{p_\nu u^\nu}{T} \right ) \pm 1} = \sum_{i=1}^{\infty} (\mp 1)^{i-1} 
\exp\left ( - i \frac{p_\nu u^\nu}{T} \right )\,  .
\end{equation}
Usually, only for pions a few terms of the expansion must be taken, otherwise the
Boltzmann approximation, {\it i.e.}, the first term, provides a good approximation. For simplicity, we will use this 
approximation here. 
Then, the $\eta$-integration can be performed explicitly and one gets
\begin{multline}
\label{e:spec1}
\frac{d^3N}{p_t\, dp_t\, d\psi\, dy} = \frac{1}{4\pi^3} \tau_0\, R_0^2 \, m_t \int_0^1 d\tilde r \, \tilde r
\int_0^{2\pi} d\phi 
\\
 \exp\left ( -\frac{p_t \cos(\psi - \phi_b) \sinh \eta_t (\tilde r,\phi_b)}{T} 
\right )
K_1\left (
\frac{m_t \cosh\eta_t (\tilde r,\phi_b)}{T}
\right )\,  ,
\end{multline}
where $K_1$ is the modified Bessel function. 
For azimuthally-integrated spectrum, the integration over azimuthal angle of the emitted particle $\psi$ leads to
\begin{equation}
\label{e:v2denomg}
\frac{d^2N}{p_t\, dp_t\, dy} = \frac{\tau_0\, R_0^2 \, m_t}{2\pi^2}  \int_0^1 d\tilde r \, \tilde r
\int_0^{2\pi} d\phi \, I_0\left (  \frac{p_t \sinh\eta_t(\tilde r, \phi_b)}{T} \right )
K_1\left (
\frac{m_t \cosh\eta_t (\tilde r,\phi_b)}{T}
\right )\,  ,
\end{equation}

As a side note, let me mention that for \emph{azimuthally symmetric sources}, there is no dependence on $\phi$ 
within the integral, so it can be performed explicitly
\begin{equation}
\frac{d^2N}{p_t\, dp_t\, dy} = \frac{\tau_0\, R_0^2 \, m_t}{\pi} \int_0^1 d\tilde r \, \tilde r
 I_0\left (  \frac{p_t \sinh\eta_t(\tilde r)}{T} \right )
K_1\left (
\frac{m_t \cosh\eta_t (\tilde r)}{T}
\right )\,  .
\end{equation}
This is the formula that is often used in fitting the single-particle $p_t$ spectra.

Let us now turn to the elliptic flow. For the sake of clarity, I repeat Eq.~(\ref{e:1}) here
\begin{equation}
\label{e:v2def}
v_2(p_t,y) = \frac{\int_0^{2\pi} \frac{d^3N}{p_t\, dp_t\, d\psi\, dy} \cos(2\psi)\, d\psi}{\int_0^{2\pi} \frac{d^3N}{p_t\, dp_t\, d\psi\, dy}\, d\psi}\, ,
\end{equation}
where we have assumed that the event plane coincides with $\psi= 0$. The denominator is given by Eq.~(\ref{e:v2denomg}). 
Adding $\cos(2\psi)$ to the integral in Eq.~(\ref{e:spec1}) and integrating over $\psi$, we obtain for the numerator
\begin{multline}
\label{e:v2numg}
\frac{\tau_0\, R_0^2 \, m_t}{4\pi^3}  \int_0^1 d\tilde r \, \tilde r
\int_0^{2\pi} d\phi \int_0^{2\pi} d\psi \, \cos(2\psi)\,
\\ \exp\left ( -\frac{p_t \cos(\psi - \phi_b) \sinh \eta_t (\tilde r,\phi_b)}{T} 
\right )
K_1\left (
\frac{m_t \cosh\eta_t (\tilde r,\phi_b)}{T}
\right )
\\
= 
\frac{\tau_0\, R_0^2 \, m_t}{2\pi^2}
 \int_0^1 d\tilde r \, \tilde r
\int_0^{2\pi} d\phi \, \cos(2\phi_b)\\
\times  I_2\left ( \frac{p_t  \sinh \eta_t (\tilde r,\phi_b)}{T} 
\right )
K_1\left (
\frac{m_t \cosh\eta_t (\tilde r,\phi_b)}{T}
\right )\,  .
\end{multline}
These formulae will be developed further, once we fully specify the azimuthally varying model.

\subsection{STAR parametrization}
\label{ss:STAR}

\begin{figure}[t]
\centerline{\includegraphics[width=0.47\textwidth]{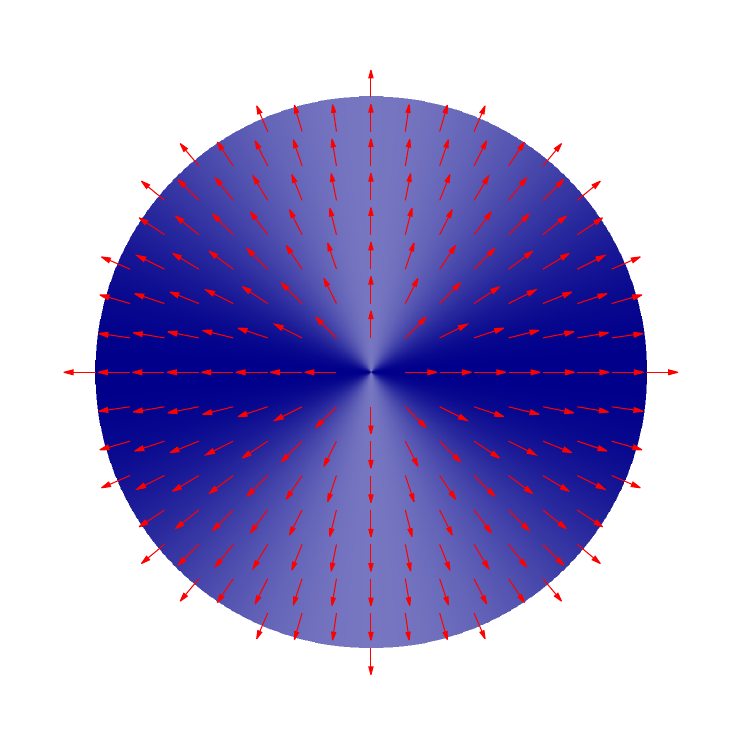}
\includegraphics[width=0.47\textwidth]{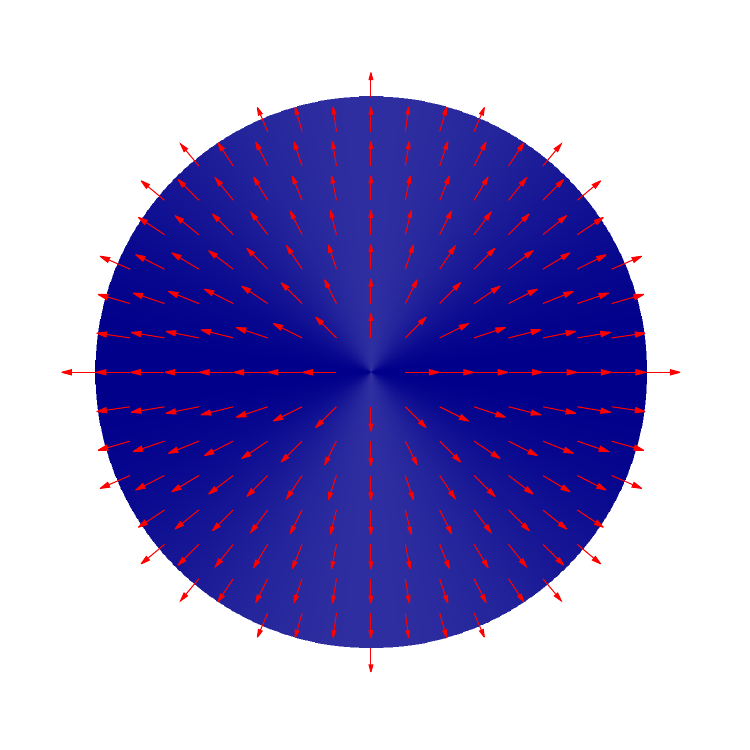}}
\caption{Schematic illustration of transverse profile and transverse flow velocity field for the parametrisation 
given by Eqs.~(\ref{e:densvar}) and (\ref{e:ravvar}). The darkness of the color reflects the source density and the arrows 
show the transverse expansion velocity. Both profiles lead to approximately same (average) elliptic flow. 
Left: larger density variation and weaker flow variation ($s_2 = 0.15$, $\rho_2 = 0.05$). Right: weaker density 
variation and stronger flow variation ($s_2 = 0.05$, $\rho_2 = 0.10$). 
\label{f:star_ill}
}
\end{figure}

Equation (\ref{e:v2def}) with numerator and denominator calculated via eqs.~(\ref{e:v2numg}) and (\ref{e:v2denomg}), respectively, 
can be used to calculate $v_2$. 
So far, only the flow field has been modified in eq.~(\ref{e:asvelo}) to exhibit  azimuthal variation. However, already in 2001  STAR collaboration realised\cite{STAR:2001ksn} 
that such a model does not fit the differential elliptic flow particularly well, but better description
can be achieved if 
the points with different velocity within the fireball receive azimuthally depending weights  in Eq.~(\ref{e:Sbase})
\begin{equation}
\label{e:densvar}
\theta(R_0-r) \to \theta(R_0-r) (1 + 2s_2 \cos(2\phi))\,  .
\end{equation} 
Such a re-weighting could be interpreted as azimuthally-dependent density profile, which will be illustrated in Fig.~\ref{f:star_ill}. A longer discussion of its interpretation is deferred to the concluding Section.

Furthermore, the calculation of $v_2$ was simplified and accelerated by replacing the $\tilde r$-dependent  transverse flow rapidity by its \emph{mean value $\bar \rho_0$}.
\begin{equation}
\label{e:ravvar}
\eta_t = \tilde r \rho_0 (1 + 2\rho_2\cos 2\phi ) \, \to \, \bar \rho_0 (1 + 2\rho_2\cos(2\phi) )\,  .
\end{equation}
Then,  the $\tilde r$-integration can be done analytically and leads to a factor 1/2. 

In the experimental papers, it is usually not described how $\bar \rho_0$ is calculated.
If one would just average the transverse rapidity without taking into account its angular variation over the circular transverse profile, 
then 
\begin{equation}
\label{e:rh0a}
\bar \rho_0 = \frac{1}{\pi} \int_0^{2\pi} d\phi \int_0^1 d\tilde r\, \tilde r \, \rho_0\tilde r = \frac{2}{3} \rho_0\, . 
\end{equation}
If the average takes into account the second-order variation of the velocity, this would change the result just by a few per cent. 
However, one could also take into account the weighting by source density, accounted for by Eq. (\ref{e:densvar}). 
Additionally, one could argue that the relevant  quantity to be averaged should be the transverse velocity instead of the rapidity and $\bar \rho_0$ should be adjusted accordingly. 
A practical possibility would also be, that $\bar \rho_0$ is just a fit parameter to reproduce $v_2(p_t)$ in the best way, but in this case, such an analysis would be disconnected from analyses of the freeze-out using  single-particle spectra because those are very sensitive to the value of $\rho_0$.
To summarize:  if the authors of an analysis do not reveal the information, we do not know \emph{what kind} of average is meant by $\bar \rho_0$.
In this note, I assume the simplest averaging given by Eq.~(\ref{e:rh0a}).

The emission function is illustrated in Fig.~\ref{f:star_ill}. 
%
%
I have deliberately made the variation not terribly huge. It is harder to see but shows a realistic picture. 
The same $v_2$ may be obtained by weaker flow variation and stronger density variation, or vice versa. 
The details, like the exact shape of the $v_2(p_t)$ dependence, may be different (see below), but the rough 
picture is this. 

Since the profile is circular rather than deformed, the angle perpendicular to the surface is just the regular azimuthal angle, $\phi_b = \phi$.
The resulting formula for $v_2$ from these adjustments is then\cite{STAR:2001ksn}
\begin{equation}
v_2 = \frac{\int_0^{2\pi} d\phi \,(1 + 2s_2 \cos(2\phi))\, \cos(2\phi)\, I_2\left ( \frac{p_t  \sinh \eta_t (\phi)}{T} 
\right )
K_1\left (
\frac{m_t \cosh\eta_t (\phi)}{T}
\right )}{%
\int_0^{2\pi} d\phi \,(1 + 2s_2 \cos(2\phi))\,  I_0\left ( \frac{p_t  \sinh \eta_t (\phi)}{T} 
\right )
K_1\left (
\frac{m_t \cosh\eta_t (\phi)}{T}
\right )
}\,  .
\label{e:star-param}
\end{equation}
The structure involving Bessel functions yields vanishing $v_2(p_t=0)$. It can be shown that for heavy particles, elliptic flow grows  quadratically with small $p_t$ (see Appendix).

\subsection{Shape anisotropy}
\label{s:sani}

The substitution (\ref{e:densvar}) may be sufficient for the elliptic flow, since $v_2$ is weakly  
sensitive to the actual shape of the source\cite{STAR:2001ksn,Retiere:2003kf,Tomasik:2004bn}. 
Nevertheless, from azimuthally sensitive femtoscopy we know that there is deformation of the fireball \emph{shape}. 
Data indicate\cite{STAR:2003ytv,STAR:2004qya,PHENIX:2014pnh,PHENIX:2015jaj,STAR:2014shf, ALICE:2017gxt}
that it is elongated rather perpendicularly to the event plane and the expansion velocity is perpendicular to the elliptic subshells of the transverse profile. 
Also, it is hard to imagine how the fireball could have reached a profile like shown in Fig.~\ref{f:star_ill}, 
if it started from a vertically elongated shape.
To describe this, the formalism was developed in Refs.~\refcite{Retiere:2003kf,Tomasik:2004bn} and in Ref.~\refcite{Cimerman:2017lmm} it was re-formulated so that it can be generalised to deformations of higher orders. 
Here we stick to this last version but stay at the second-order level. 

The transverse radius is assumed to oscillate with the azimuthal angle $\phi$
\begin{equation}
R(\phi) = R_0 ( 1 - a_2 \cos 2\phi)\, ,
\end{equation}
where the sign  is chosen so that positive $a_2$ leads to positive $v_2$.

The dimensionless radial coordinate is then re-defined
\begin{equation}
\tilde r \to \tilde r = \frac{r}{R(\phi)}
\end{equation}
so that in the integral, the variable is changed 
\begin{equation}
r\, dr = \tilde r\, d\tilde r\, R_0^2\, \to \, \tilde r\, d\tilde r\, R_0^2 (1-a_2 \cos2\phi)^2\, .
\end{equation}
Since we have a fireball with truly deformed shape now, also the angle $\phi_b$ perpendicular to the surface becomes different from $\phi$.
It is determined as 
\begin{equation}
\phi_b = \alpha \pm \frac{\pi}{2}\, ,
\end{equation}
where the upper (lower) sign applies for $0<\phi\le\pi$ ($\pi<\phi\le2\pi$). The angle $\alpha$ is calculated from
\begin{equation}
\tan \alpha = \frac{\frac{dy}{d\phi}}{\frac{dx}{d\phi}} = 
-\frac{\cos \phi }{\sin\phi} \frac{1+5a_2 - 6a_2\cos^2\phi}{1-5a_2 + 6a_2\sin^2\phi } \, ,
\end{equation}
where $x,y$ (only here) are the coordinates of a point along a surface of the fireball given by constant $\tilde r$.

After all these substitutions into Eqs.~(\ref{e:v2def}), (\ref{e:v2denomg}), and (\ref{e:v2numg}) one derives 
for the elliptic flow coefficient
\begin{equation}
\label{e:shapev2}
v_2 = \frac{
\int_0^1 d\tilde r\, \tilde r \int_0^{2\pi} d\phi\, (1-a_2\cos2\phi)^2 \cos(2\phi_b) \,
I_2\!\left (
\frac{p_t  \sinh \eta_t (\tilde r,\phi_b)}{T}
\right )
K_1\!\left (
\frac{m_t  \cosh \eta_t (\tilde r,\phi_b)}{T}
\right )
}{
\int_0^1 d\tilde r\, \tilde r \int_0^{2\pi} d\phi \, (1-a_2\cos2\phi)^2 
I_0\!\left (
\frac{p_t  \sinh \eta_t (\tilde r,\phi_b)}{T}
\right )
K_1\!\left (
\frac{m_t  \cosh \eta_t (\tilde r,\phi_b)}{T}
\right )
}\, .
\end{equation}
Notice that both $\phi$ and $\phi_b=\phi_b(\phi)$ appear in the integrals.

\begin{figure}[t]
\centerline{\includegraphics[width=0.47\textwidth]{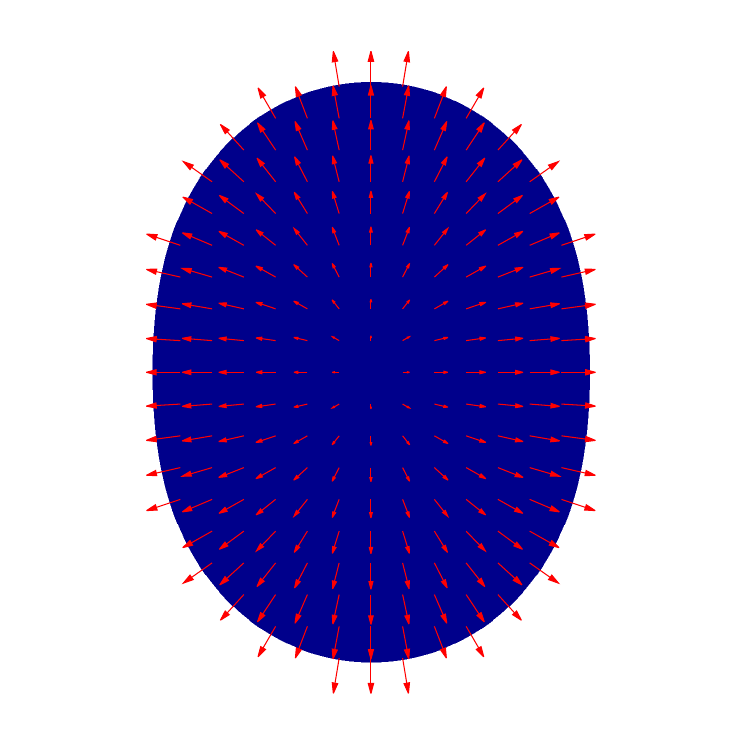}
\includegraphics[width=0.47\textwidth]{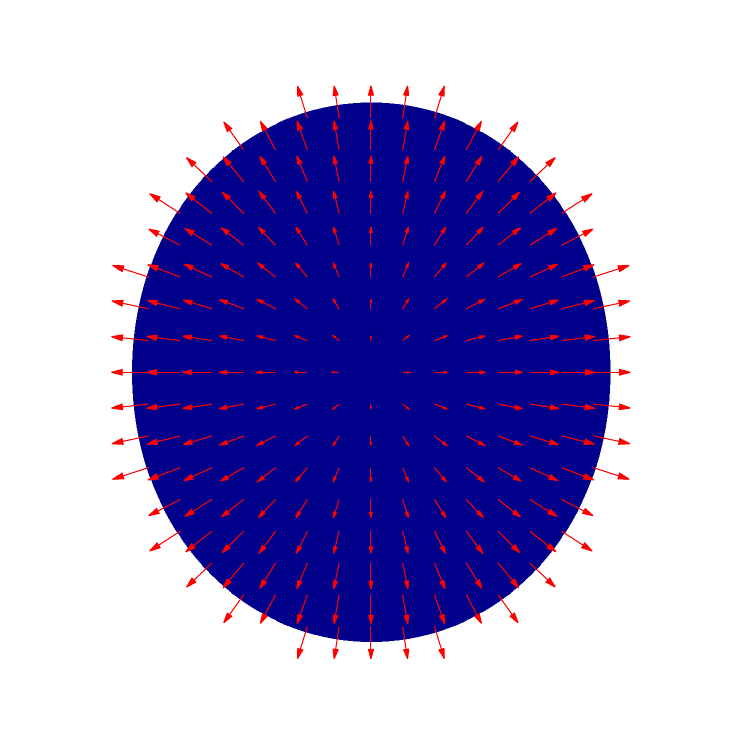}}
\caption{Schematic illustration of transverse profile and transverse flow velocity field for the shape anisotropy parametrisation 
described in Sec.~\ref{s:sani}.
Left: larger spatial anisotropy and weaker flow variation ($a_2 = 0.14$, $\rho_2 = 0.015$). Right: weaker 
spatial anisotropy and stronger flow variation ($s_2 = 0.06$, $\rho_2 = 0.055$). 
\label{f:myprof}
}
\end{figure}

The model is illustrated in Fig.~\ref{f:myprof}.

There is a qualitative difference between this model and the STAR parametrization from Section \ref{ss:STAR}.
Here, the formulation of the fireball spatial geometry is motivated by femtoscopy, while the STAR parametrization does not take femtoscopy into account: density profiles illustrated in Fig,~\ref{f:star_ill} are certainly unrealistic, but the aim in setting up that model was just to reweight the emission points and not really address the  spatial distribution of the fireball reflected in femtoscopy. 
I will discuss this issue again in Conclusions.

\section{Elliptic flow phenomenology with the models}
\label{s:exams}

As I have already indicated, the anisotropy of hadron distribution---quantified by $v_2$---may result from an anisotropy of the expansion flow velocity but also from an anisotropy of source shape or density distribution.
To illustrate this ambiguity, I fix the temperature to 106~MeV and the radial flow gradient $\rho_0 = 0.91$, or 
$\bar \rho_0 = 0.6$ for the STAR parametrisation. 
The choice of parameters is guided by data from Pb+Pb collisions at $\sqrt{s_{NN}} = 2.76$~TeV, centrality class 30-40\%, 
as measured by the ALICE collaboration \cite{Melo:2019mpn,ALICE:2013mez}.
The parameters of the azimuthal variation for all calculations which follow are summarized in Table~\ref{t:params}.
%
\begin{table}
\tbl{Anisotropy parameters used in the illustrative calculations.}
{\begin{tabular}{clcc}
\hline
 \# & parameter set name  & $\rho_2$ & $a_2$ or $s_2$ \\
 \hline
 \hline
 \multicolumn{4}{c}{Examples of STAR parametrisation}\\
 \hline
1 & best fit to data & 0.10300 & 0.05690 \\
2 &  space anisotropy dominance & 0.07500 & 0.09000\\
3 &  flow anisotropy dominance & 0.12500 & 0.02000\\
\hline\hline
\multicolumn{4}{c}{Examples of shape anisotropy}\\
\hline
4 & best fit to data & 0.04075 & 0.08400\\ 
5 & shape anisotropy dominance for pions &0.02800& 0.14000 \\
6 & shape anisotropy dominance for protons  &0.02800&0.03800\\
7 & flow anisotropy dominance for pions  &0.05600&0.10800\\
8 & flow anisotropy dominance for protons  &0.05600&0.06200\\
\hline\hline
\multicolumn{4}{c}{Comparison STAR vs shape anisotropy}\\
\hline
9 & best agreement for pions & 0.104320 & 0.06048 \\
10 & best agreement for protons & 0.087205 & 0.05712 \\
11 & best agreement overall & 0.104320 & 0.05208 \\
\hline\hline
\multicolumn{4}{c}{Fits to data, STAR parametrisation}\\
\hline
12 & best fit to pions & 0.06815 & 0.11550\\   
13 & best fit to protons  & 0.07700 &  0.04300 \\
14 & best fit overall   &   0.10300  &  0.05690\\
\hline\hline
\multicolumn{4}{c}{Fits to data, shape anisotropy}\\
\hline  
15 & best fit to pions & 0.02425 & 0.15250 \\
16 & best fit to protons &  0.03700 & 0.06500\\
17 & best fit overall & 0.04075 &0.08400 \\
\hline
\end{tabular}\label{t:params}}
\end{table}

To explore the similarity of the calculated $v_2(p_t)$ for different model parameters, I first calculate the reference results, $v_{2,\mathrm{ref}}(p_t)$.
For this, I will usually choose model parameters that are best in reproducing experimental data (with caveats to be discussed later). Then, I will vary anisotropy parameters $\rho_2$ and $a_2$ or $s_2$, and measure how much different the resulting $v_2(p_t)$ is from the reference.
The difference is quantified by 
\begin{equation}
\label{e:delta}
\Delta = \int_0^{p_m} (v_{2,\mathrm{ref}}(p_t) - v_2(p_t))^2 dp_t\, ,
\end{equation}
where $p_t$ is measured in GeV and I choose $p_m = 1.5$~GeV.  The formula will be applied on results for identified pions or identified protons. 
Since this is a difference of two theoretical curves, the choice of an appropriate measure is arbitrary, and I will not draw conclusions from the actual size of $\Delta$, but use it just in relative sense and in order to see which parameters yield similar results and which are totally different.

\begin{figure}[t]
\centerline{\includegraphics[width=0.58\textwidth]{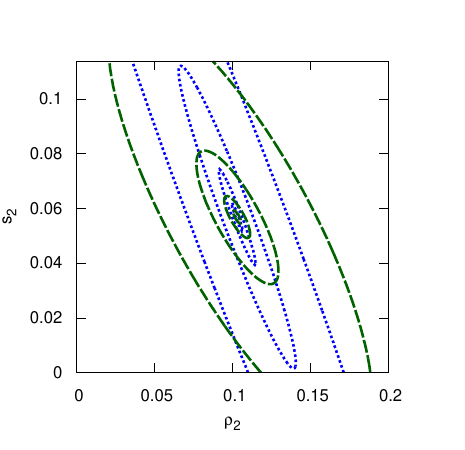}
\includegraphics[width=0.41\textwidth]{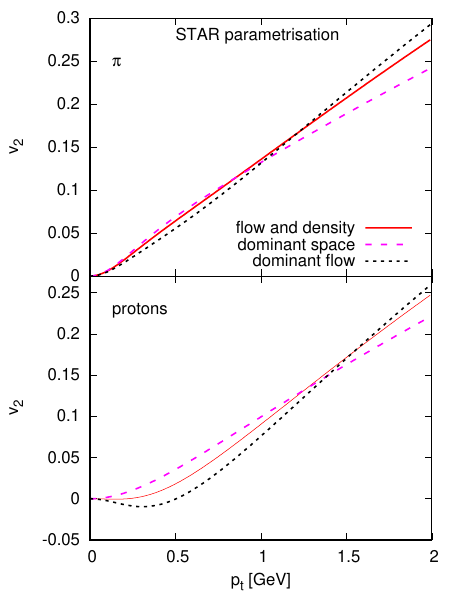}}
\caption{Left: Parameter map for the STAR parametrisation (Eq.~(\ref{e:star-param})). Reference calculation with ``flow and density'' variation done with set \#1 from Table~\ref{t:params}. Contours indicate difference $\Delta$ from Eq.~(\ref{e:delta}) equal to $10^{-6}$, $10^{-5}$, $10^{-4}$, and $10^{-3}$.
Blue dotted contours are for the difference in pion $v_2(p_t)$ and green dashed contours are for the difference in proton $v_2(p_t)$. Right: The $v_2(p_t)$ 
dependences calculated for pions (upper panel) and protons (lower panel) for parameters of the reference set, dominant spatial density variation (set \#2 in Table~\ref{t:params}) and dominant flow variation (set \#3). The parameter sets are chosen from within the $10^{-4}$ contour in the left panel. 
\label{f:mapstar}
}
\end{figure}

The results from this exercise for the STAR parameterization are summarized in Fig.~\ref{f:mapstar}.
Recall that we want to check the ambiguity in determining the flow anisotropy $\rho_2$ and the density anisotropy $s_2$ from the measured $v_2(p_t)$. 
If looking at a single hadron species only, this is indeed the case.  
The most pronounced feature is the anti-correlation between $\rho_2$ and $s_2$. 
Thus, the flow anisotropy can be compensated by density anisotropy or vice versa, to get the same $v_2(p_t)$.
However, the figure also shows that the anti-correlation is different for pions, and for protons, i.e., it depends on the mass of the hadrons.
The $v_2(p_t)$ dependences were calculated with parameters along the valley that indicates good agreement between the results.
The differences between the results are particularly visible for protons.
Flow anisotropy leads to lower $v_2$ at small $p_t$, which then grows faster as $p_t$ increases.
The anisotropy of the density profile causes $v_2(p_t)$ that increases immediately at low $p_t$ and slows down the growth as $p_t$ increases above 1 GeV.
It is suggestive  that such a sensitivity to hadron mass and both kinds of anisotropies would allow to determine from data both $\rho_2$ and $s_2$ uniquely.
Nevertheless, one should not forget that the model is rather schematic and should not be interpreted too literally.
I shall  discuss this further below.

\begin{figure}[t]
\centerline{\includegraphics[width=0.58\textwidth]{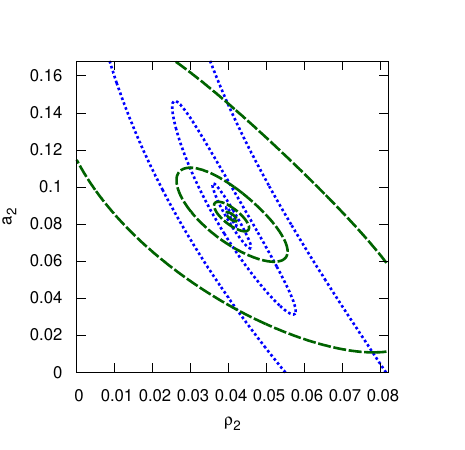}
\includegraphics[width=0.41\textwidth]{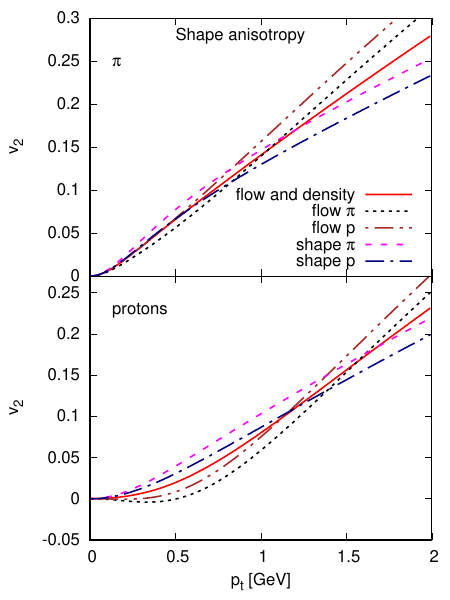}}
\caption{Left: Parameter map for the parametrisation with shape anisotropy (Eq.~(\ref{e:shapev2})). Reference calculation with ``flow and density'' variation done with set \#4 from Table~\ref{t:params}. Contours indicate difference $\Delta$  equal to $10^{-6}$, $10^{-5}$, $10^{-4}$, and $10^{-3}$.
Blue dotted contours are for the difference in pion $v_2(p_t)$ and green dashed contours are for the difference in proton $v_2(p_t)$. Right: The $v_2(p_t)$ 
dependences calculated for pions (upper panel) and protons (lower panel) for parameters of the reference set. 
They are compared with results with dominant flow anisotropy that fit better the pion $v_2(v_2)$ (set \#7, black dotted curve), flow anisotropy that better fits the protons (set \#8, brown dash-dot-dot), dominant shape anisotropy that fits pions (set \#5, magenta dashed) and shape anisotropy fitting protons (set \#6, dark-blue, dash-dot).    The sets are chosen within the $10^{-4}$ contour of pions or protons in the left panel.
\label{f:mapbw}
}
\end{figure}

The analogical results from the shape-anisotropy model are shown in Fig.~\ref{f:mapbw}.
%
Again, an anti-correlation between flow and shape anisotropy is seen in the left panel, but this time the difference 
between pions and protons is much larger. 
In order to get similar $v_2(p_t)$, much weaker variation of shape anisotropy $a_2$ is allowed for protons than for pions. 
Due to the difference between protons and pions I calculated $v_2(p_t)$ for parameters that lead to similar pion $v_2(p_t)$ and separately for parameters that gain it similar for protons. 
We see that the initial decrease of $v_2(p_t)$ for small $p_t$ with protons is not due to $\rho_2$ only, but is enhanced 
by larger $a_2$, as well.

\begin{figure}[t]
\centerline{\includegraphics[width=0.58\textwidth]{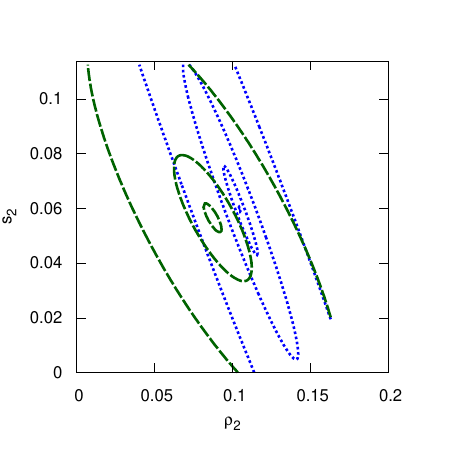}
\includegraphics[width=0.41\textwidth]{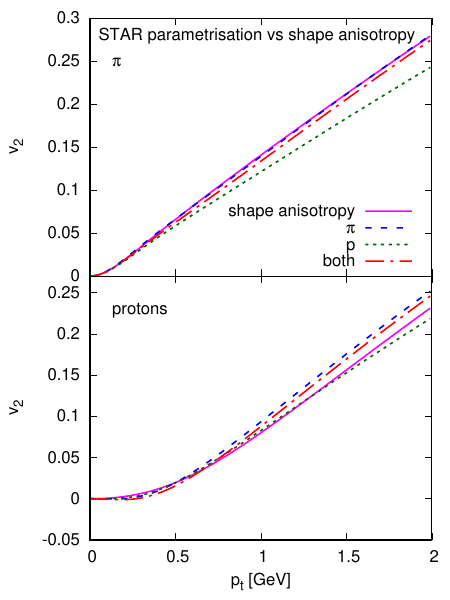}}
\caption{Similar to Figs.~\ref{f:mapstar} and \ref{f:mapbw}, but calculations with STAR parametrisation are compared to 
reference results from the shape-anisotropy model (set \#4). Right: reference results (set \#4) compared with 
$v_2(p_t)$ from STAR parametrisation. Blue dashed line: best agreement with pions (set \#9); green dotted: best agreement with protons (set \#10); red dash-dotted: best overall agreement with both pions and protons (set \#11).  
\label{f:mapcomp}
}
\end{figure}

It is quite natural to compare the two models.
Since the shape-anisotropy model is more realistic, I again calculated reference $v_2(p_t)$ with this model (set \#4) 
and then compared the results of the STAR parameterization to it.
The map of $\Delta(\rho_2,s_2)$  contours and $v_2(p_t)$ are plotted in Fig.~\ref{f:mapcomp}.
%
A brief inspection of Eqs.~(\ref{e:star-param}) and (\ref{e:shapev2}) could lead to an idea that the results might be similar if $\rho_2$ is kept the same and $s_2=-a_2$.
The reasoning would be, that for small $a_2$ we have $(1-a_2\cos(2\phi))^2 \approx (1 - 2a_2 \cos(2\phi))$
and the formula would be similar.
The results show that this is not true, at all: the flow anisotropy $\rho_2$ required in STAR parameterization is more than twice as big as in the shape-anisotropy model, and $s_2$ is of the same sign and somewhat smaller than $a_2$, as can be seen in Table~\ref{t:params}.
On the quantitative level, the two models yield completely different results.
Thinking more carefully one realises that there are still many differences between the models. 
Particularly, the direction of the flow $\phi_b$, which boosts the emission of hadrons, differs from $\phi$ in the shape-anisotropy extension of the blast-wave model.

\begin{figure}[t]
\centerline{\includegraphics[width=0.47\textwidth]{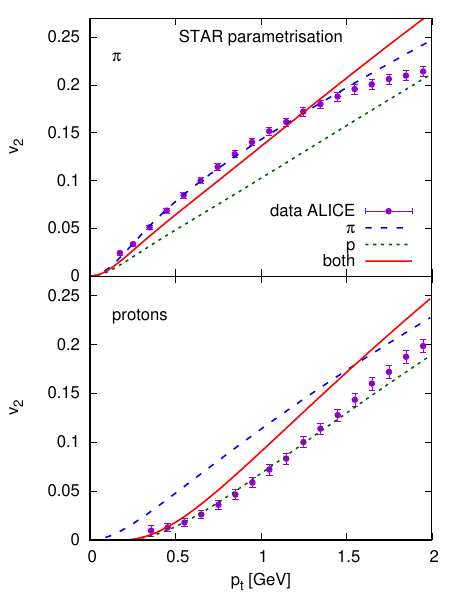}
\includegraphics[width=0.47\textwidth]{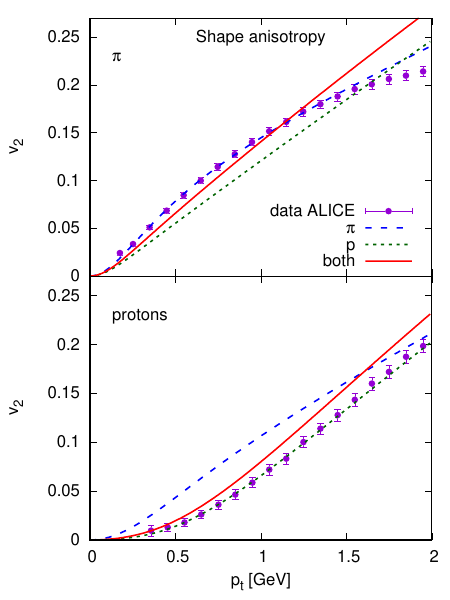}}
\caption{Theoretical curves compared to data on $v_2(p_t)$ of identified pions and protons from Pb+Pb collisions at $\sqrt{s_{NN}} = 2.76$~TeV, centrality class 30-40\% \cite{ALICE:2013mez}. Calculations are performed with parameters from best fits 
to pions (blue dashed curves), protons (green dotted), and both pions and protons (red solid). Left: calculations with 
STAR parametrisation, Eq.~(\ref{e:star-param}); right: shape anisotropy model, Eq.~(\ref{e:shapev2}). 
\label{f:fitexams}
}
\end{figure}

This is even more visible when the models are compared to data.
Just for the sake of an illustration I choose dataset  from Pb+Pb collisions at $\sqrt{s_{NN}} = 2.76$~TeV, 
centrality 30-40\%\cite{ALICE:2013mez}.
\begin{figure}[t]
\centerline{\includegraphics[width=0.7\textwidth]{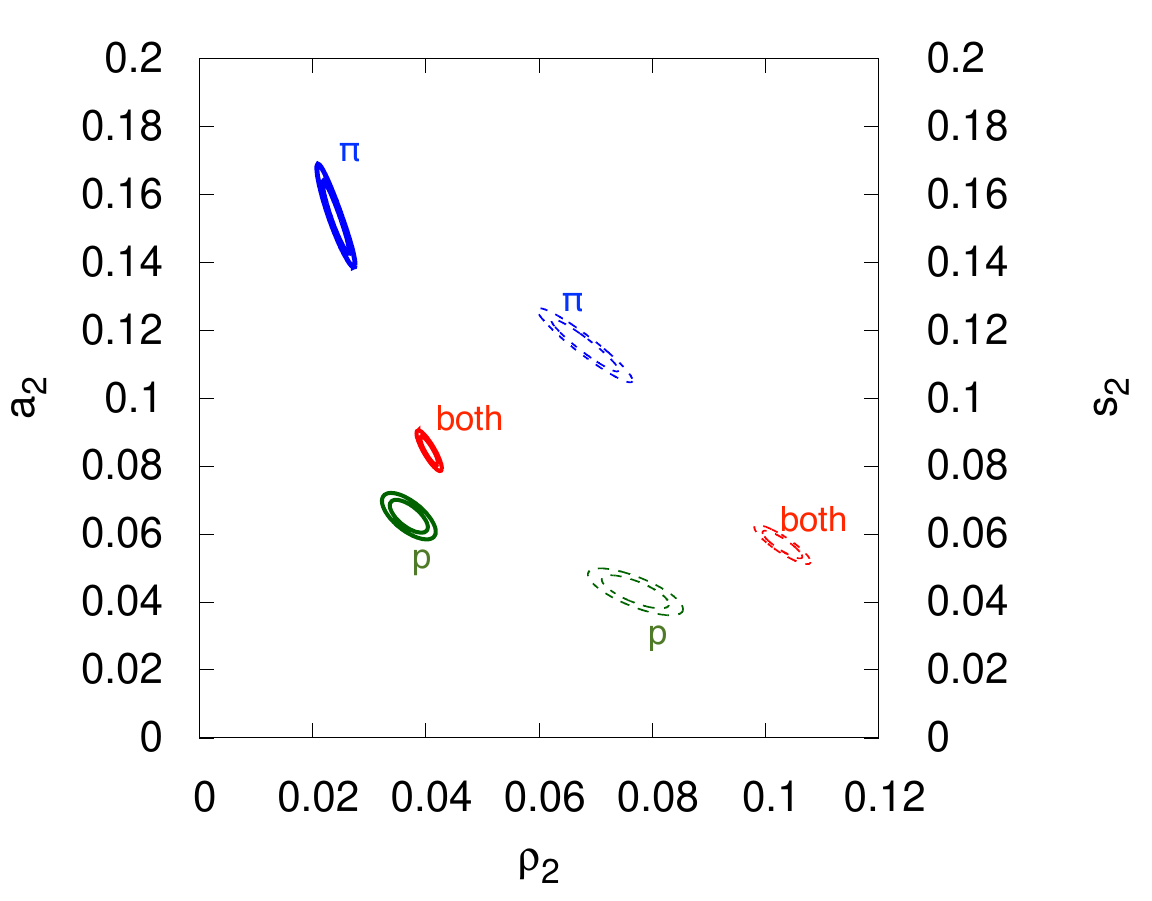}}
\caption{Results from fitting $v_2(p_t)$ of identified pions and protons from Pb+Pb collisions at $\sqrt{s_{NN}} = 2.76$~TeV, centrality class 30-40\% \cite{ALICE:2013mez}. Contours of $1\sigma$ and $2\sigma$ around best fits are shown. Solid lines: shape-anisotropy (best fits sets \#15, \#16, \#17); dashed lines: STAR parametrisation (best fits sets \#12, \#13, \#14). Blue: fits to pions; green: fits to protons; red: fits to pions and protons simultaneously.  
\label{f:fitcontours}
}
\end{figure}
In Fig.~\ref{f:fitcontours} I show the $1\sigma$ and $2\sigma$ contours in the parameter space for both models.
Clearly, best fits require about twice as large $\rho_2$ for STAR parameterization, than for the shape anisotropy.
It is also seen that pions and protons require mutually incompatible sets of parameters for the best fit.
A best fit to  both pions and protons  can be identified, but actually reproduces neither of the two measured $v_2(p_t)$, as can be seen in Fig.~\ref{f:fitexams}.
This holds for both models. Hence, by trying to fit $v_2(p_t)$ of identified hadrons we may either determine the parameter set uniquely \cite{Cimerman:2017lmm}  or---like here---exclude the models \cite{Vozabova:2024jum}.
I will comment on this situation further below.

\section{Conclusions}
\label{s:conc}

The example results that I have shown carry some general messages.
Before I start with them, let me recall that the temperature and radial transverse flow gradient determine the azimuthally integrated transverse momentum spectra.
Those spectra depend only very weakly  on the anisotropies.
Hence, before studying $v_2(p_t)$, temperature and $\rho_0$ can be set from the spectra, and later one can only focus on the anisotropy parameters.

Firstly, precise experimental data on the $v_2(p_t)$ dependence
indeed seem to convey some information about the exact way how flow and shape or density anisotropy appear in the fireball.
If only one sort of identified hadrons is analyzed, this leads to unique determination of the model parameters.
However, for the two models explored here, protons and pions were never reproduced by the same parameter set.
This forces us to conclude that these models do not describe the final state of the fireball in non-central collisions.
They do not even provide any reasonable \emph{effective} description, since each particle species requires different parameter set.
To put the moral shortly: Never interpret such fit results to identified $v_2(p_t)$ as the true picture of the fireball, if the fit does not the reproduce different particle species.

Secondly, I also explained that there is no well established and broadly accepted blast-wave model for non-central collisions, 
even though in many conference talks you will encounter ``the blast-wave results''  for the elliptic flow.
I have motivated and formulated here two possible extensions of the established and standard azimuthally symmetric blast-wave model, one proposed by the STAR collaboration and the other that gradually evolved from Refs.~\refcite{Retiere:2003kf,Tomasik:2004bn,Cimerman:2017lmm}.

Unfortunately, there is no quantitative correspondence between them. In other words, the $v_2(p_t)$ calculated in these models differ, and the anisotropy parameter values required to reproduce similar $v_2(p_t)$ are different and cannot be transferred from one model to the other.
This argues against using the values of the anisotropy parameters in any context outside the framework of a particular model.

Recall also that the STAR parametrization was probably motivated to a large extent just by economy of computation, but the profile shown in Fig.~\ref{f:star_ill} is not realistic as it would not fit the azimuthally sensitive femtoscopic radii.
The shape-anisotropy parametrization was developed with femtoscopy results in mind \cite{Retiere:2003kf,Tomasik:2004bn,Cimerman:2017lmm}, but it does not fit the $v_2(p_t)$ here either---read the next point.

Thirdly, after trying to fit identified $v_2(p_t)$ for just one centrality class from Pb+Pb collisions at $\sqrt{s_{NN}}= 2.76$~TeV, we have to conclude that none of the models really describes experimental data for protons and pions.
None of the two extensions of the blast-wave model that were formulated here really describes the state of the fireball at the moment of its kinetic break-up.

I actually encountered this problem when we were looking for the parametrization of the fireball that would best fit the data, 
because we wanted to use it for a study of deuteron production \cite{Vozabova:2024jum}.
We wanted to describe the freeze-out state but did not want to go to extensive simulations of the whole fireball evolution.
We eventually did manage to simultaneously describe $v_2(p_t)$ of pions and protons, however, with a much more developed model than was presented here.
Although based on the shape-anisotropy generalization shown above, our model there also involved
i) modified freeze-out hypersurface with freeze-out time depending on $r$; 
ii) resonance production and decays that also produced pions and/or protons;
iii) corrections to the statistical distribution of hadron momenta due to viscosity of the fireball matter \cite{McNelis:2021acu}.

There appears sometimes an argument, that unless one is interested in the actual spatial distribution of the emitting source as measured by femtoscopy, then for the interpretation of $v_2$  only points with different velocities within the source need to be weighted appropriately, and how the fluid velocity depends on spatial coordinates is of little relevance. 
The STAR parametrization is an example of such an approach.
Let us look at the argument.
Hadrons with given momentum are produced not just from one point but from a whole homogeneity region, which consists of all places that move with velocities not too much different from those of the hadrons in consideration. 
Thus it is important how the individual emission points are assembled into homogeneity regions and how those regions are weighted.
Mathematically, it would be enough just to choose these weights properly and $v_2$ should be reproduced. 
Nevertheless, it is better to aim at a model that potentially can address more than just a single observable. 
A correct distribution in space-time should also reproduce femtoscopy and the weights of different homogeneity regions that are relevant for $v_2$ then follow from it.

Finally, one could always argue that not being able to fit $v_2(p_t)$ data with a \emph{parametrization} is not a problem, 
since what really matters is whether we have a \emph{simulation} that is able to reproduce all the details.
One could completely drop the idea of fitting data with the blast-wave model due to such reasoning.
However, a parametrization like the blast-wave model may provide an inexpensive description of the reality that could help to gain better insight into the evolution of the fireball.
Therefore, the motivation to find an azimuthally dependent extension of the blast-wave model that describes the data is still pertinent, even in the era of sophisticated hybrid models and involved Bayesian data analyses.


\section*{Acknowledgments}

This research was supported  by the Czech Science Foundation under No. 22-25026S. 
The participation to Zim\'anyi School was covered by VEGA 1/0521/22.


\appendix

\section{Elliptic flow at small momentum}

In this appendix, expressions for $v_2(p_t)$ are derived for the limiting cases of very small transverse momentum, based on formulas (\ref{e:star-param}) and/or (\ref{e:shapev2}). 
For small transverse momentum, we can use the expansion formulae for modified Bessel functions valid for small argument $x$
\begin{eqnarray}
I_0(x) & \approx & 1 + \frac{x^2}{2} \\
I_2(x) & \approx & \frac{x^2}{8} + \frac{x^4}{64} \\
K_1(x) & \approx & \frac{1}{x} \, .
\end{eqnarray}
In both formulas (\ref{e:star-param}) and (\ref{e:shapev2}), the arguments of the $I_\nu$ Bessel functions are proportional to $p_t$, while the argument of $K_1$ involves $m_t$. For large mass of the particle, one can neglect $p_t$ against the mass, $m_t = \sqrt{m^2 + p_t^2}\to m$, and we obtain for the STAR parametrisation up to leading order in $p_t$
\begin{multline}
v_2 = \frac{\int_0^{2\pi} d\phi \,(1 + 2s_2 \cos(2\phi))\, \cos(2\phi)\,  \frac{1}{8}\frac{p_t^2   \sinh^2\! \eta_t (\phi)}{T^2}\,
K_1\left (
\frac{m\cosh\eta_t (\phi)}{T}
\right )}{%
\int_0^{2\pi} d\phi \,(1 + 2s_2 \cos(2\phi))\,  
K_1\left (
\frac{m \cosh\eta_t (\phi)}{T}
\right )
}
\\
=
\frac{1}{8}\frac{p_t^2}{T^2} \frac{\int_0^{2\pi} d\phi \,(1 + 2s_2 \cos(2\phi))\, \cos(2\phi)\,   \sinh^2\! \eta_t (\phi)\,
K_1\left (
\frac{m \cosh\eta_t (\phi)}{T}
\right )}{%
\int_0^{2\pi} d\phi \,(1 + 2s_2 \cos(2\phi))\,  
K_1\left (
\frac{m \cosh\eta_t (\phi)}{T}
\right )
}\, .
\end{multline}
For the shape anisotropy model one obtains
\begin{multline}
v_2 = \frac{%
\int_0^1 d\tilde r\, \tilde r \int_0^{2\pi} d\phi\, (1-a_2\cos2\phi)^2 \cos(2\phi_b) \,
\frac{1}{8}
\frac{p_t^2  \sinh^2\! \eta_t (\tilde r,\phi_b)}{T^2}
\,
K_1\!\left (
\frac{m  \cosh \eta_t (\tilde r,\phi_b)}{T}
\right )
}{
\int_0^1 d\tilde r\, \tilde r \int_0^{2\pi} d\phi \, (1-a_2\cos2\phi)^2 
K_1\!\left (
\frac{m  \cosh \eta_t (\tilde r,\phi_b)}{T}
\right )}
\\ = 
\frac{1}{8} \frac{p_t^2}{T^2} 
\\ \times
\frac{%
\int_0^1 d\tilde r\, \tilde r \int_0^{2\pi} d\phi\, (1-a_2\cos2\phi)^2 \cos(2\phi_b) \,
\sinh^2\! \eta_t (\tilde r,\phi_b)
\,
K_1\!\left (
\frac{m  \cosh \eta_t (\tilde r,\phi_b)}{T}
\right )
}{
\int_0^1 d\tilde r\, \tilde r \int_0^{2\pi} d\phi \, (1-a_2\cos2\phi)^2 
K_1\!\left (
\frac{m  \cosh \eta_t (\tilde r,\phi_b)}{T}
\right )}\,  ,
\end{multline}
hence in both cases the initial growth is quadratic in $p_t/T$. 

The situation seems similar for vanishing  mass. One has to be careful, however. 
The only relevant massless particles are photons, which are emitted from the hot fireball throughout its whole evolution. 
Therefore, the Cooper-Frye formula (\ref{e:CF}) cannot be applied, since it assumes instantaneous freeze-out along specified hypersurface.



\end{document}